\documentclass[12pt,a4paper]{article}
\usepackage{graphicx}
\usepackage{amsmath}

\begin{document}

\thispagestyle{empty}

\title{Supersymmetry, replica and dynamic treatments of disordered
  systems: a parallel presentation.}

\author{Jorge Kurchan}

\maketitle

 Running title: Supersymmetry, replicas and dynamics


 82C31 37A50 37N40

\begin{abstract}
I briefly review the three nonperturbative
methods for the treatment of disordered
systems ---
supersymmetry, replicas and dynamics --- with a parallel presentation 
that highlights their connections and differences.  

\end{abstract}
\vspace{.5cm}

Disordered systems  need to be treated with a method that allows
to perform  averages over the sample realisation.
 There is no universal way 
to do this  that can be applied efficiently to all problems. 

For Gaussian systems, 
the method of supersymmetry  
is as good as one can expect: it involves
a minimum of variables, it is elegant and rigorous. 
Although one can still apply it for some non-Gaussian problems, 
 in many of the interesting cases -- as for example  spin-glasses -- 
it only  gives  limited information.

The replica trick was introduced to tackle such `complex' problems. 
It has been extensively used and has yielded some of the most
innovative solutions in disordered systems.
 It has however the
problem that it is very far from being controlled, let alone rigorous.
This is because the space itself -- a vector space with noninteger
dimension --  
does not have a general definition other than the  ansatz itself - 
or perturbations
around it.

The dynamic method 
consists of solving exactly  the evolution of the
system in contact with a heat bath.  If the system 
reaches equilibrium one recovers all the thermodynamic information.
Surprisingly enough, one can treat this way all the problems one can
solve with replicas.
A problem arises, however, when equilibrium cannot be achieved: then the
long-time out
of equilibrium regime may be of interest in itself (as in the case of
 glasses), or it may be viewed as an obstacle for exploring the deepest levels
in phase-space (as for example in optimisation problems).
Although  the dynamic method was initially proposed
as a way to obtain equilibrium results, this tendency has reverted in
the last few years, at least in the field of glasses,
where replicas are now used mostly  to
mimic  the out of equilibrium  dynamics.

The aim is of this paper is not to make a complete presentation of 
either of the three methods ---
there are very complete reviews  of this
\cite{Efetov,Mepavi,Bocukume} (including some very recent ones \cite{Ka})
but rather to put the three methods `side by side' so that the 
connections can be better appreciated. To the best of my knowledge
this has not been done for supersymmetry, replicas and dynamics
simultaneously, as the practitioners of each method
 tend to belong to  different communities.

\vspace{.5cm}

{\bf The Problem}

\vspace{.5cm}

Consider an energy
\begin{equation}
E_{\boldsymbol{J}}= \frac{i}{2} \sum_{ij} (\lambda \delta_{ij}-J_{ij} ) s_i s_j
\;\;\;
; \;\;\;
E_{\boldsymbol{J}}(h)=E_{\boldsymbol{J}} - \sum_i h_i s_i 
\label{linear}
\end{equation}
where $s_i$ ($i=1,...,N$) are  real variables, and $J_{ij}$ is
a random matrix. We take $\lambda$ with negative
 imaginary part. 
This energy can be used to calculate the averaged 
Green function:
\begin{equation}
{\overline{G(\lambda)}} \equiv {\overline{{
\mbox{Tr}} [\lambda {\boldsymbol{I}} - 
{\boldsymbol{J}}]^{-1}
}}
\end{equation}
from which one obtains the eigenvalue distribution.
(Here and in what follows the overline denotes averages over the
disorder ${\boldsymbol{J}}$). This is done by defining the partition
function
\begin{equation}
Z_{\boldsymbol{J}}(h) = \int 
\boldsymbol{ds} \; e^{-\beta E_{\boldsymbol{J}}(h)}
\label{partition}
\end{equation}
and  computing:
\begin{eqnarray}
{\overline{
G(\lambda)
}} &=& -iT \left. 
\sum_k \frac{\partial^2}{\partial h_k^2} 
{\overline{ \ln Z_{\boldsymbol{J}}}}(h)\right|_{h_i=0} =
\left. i \beta \sum_k
{\overline{
\left[\int  \boldsymbol{ds}
 \;  s^2_k \; e^{-\beta E_{\boldsymbol{J}} (h)} \right]
\cdot \frac{1}{Z_{\boldsymbol{J}}(h)} }}\right|_{h_i=0} \nonumber\\
&{\boldsymbol{\neq}}&
\left. i \beta \sum_k
{\overline{
\left[\int  \boldsymbol{ds}
 \;  s^2_k \; e^{-\beta E_{\boldsymbol{J}} (h)} \right]}}
\cdot 
\frac{1}{{\overline{Z_{\boldsymbol{J}}(h)} }}\right|_{h_i=0}
\label{quenched}
\end{eqnarray}
(We include
 the constant normalisations in the differential:
  ${\boldsymbol{d}}\equiv d/\sqrt(2\pi N)$).  

The third expression in (\ref{quenched}) 
is the correct (quenched) average, in general different from
 the last one, the annealed average.
{\em The problem is that in order to compute the average over the
$J_{ij}$, we need to express $1/Z_{\boldsymbol{J}}$ in (\ref{quenched}) in 
a tractable (i.e. exponential)
form}. Three methods to do so are:

\begin{itemize}

\item {\em Supersymmetry}: we can take advantage of the Gaussian nature of
  the partition function to write, in terms of two sets of Grassmann 
 $\eta_i$ and $\eta_i^*$ and a set of ordinary variables $\sigma_i$:
\begin{equation}
\frac{1}{Z_{\boldsymbol{J}}} = \int  {d \boldsymbol{\eta}} \; 
{d \boldsymbol{\eta}}^*
{\boldsymbol{d\sigma}}
 \; 
e^{
-\frac{\beta}{2} i \sum_{ij} 
(\lambda \delta_{ij}- J_{ij}) (\eta_i^* \eta_j+
\sigma_i \sigma_j)  }
\label{partition-1}
\end{equation}
so that we get:
\begin{equation}
{\overline{G(\lambda)}} =
\left. \sum_k
{\overline{
\int  
{\boldsymbol{d s}} 
{\boldsymbol{d \sigma}}
\int   \boldsymbol{d \eta} \; {\boldsymbol{d \eta^*}}
 \;  s^2_k \; e^{
-\frac{\beta}{2} i \sum_{ij} 
(\lambda \delta_{ij}- J_{ij}) 
(\eta_i^*   
\eta_j + \sigma_i \sigma_j + s_i s_j )} 
}}
\right|_{h_i=0} 
\label{susy}
\end{equation}

\item {\em Replicas}:  we replicate $n$ times each variable $s_i \rightarrow
  s_i^\alpha$  and compute:
\begin{equation}
Z_{\boldsymbol{J}}^{n-1} =  \int \Pi_{\alpha=1}^{n-1} 
{\boldsymbol{d s^\alpha }} \; e^{
-\frac{\beta}{2} i \sum_{\alpha=1}^{n-1}
  \sum_{ij} (J_{ij}-\lambda \delta_{ij}) s_i^{\alpha} 
s_j^{\alpha} }
\label{rep1}
\end{equation}
The calculation proceeds for every integer $n$, and finally we
somehow take the limit:
\begin{equation}
Z_{\boldsymbol{J}}^{-1} = \lim_{n \rightarrow 0} Z_{\boldsymbol{J}}^{n-1}
\label{rep2}
 \end{equation}
which should in principle be shown to be the correct
 analytic continuation over $n$. We hence have:
\begin{equation}
{\overline{G(\lambda)}} 
\sim_{n \rightarrow 0} {\overline{
 \sum_k \int \Pi_{\alpha=1}^{n} 
d{\boldsymbol{s}}^\alpha 
\; (s_k^{(1)})^2
 \; e^{
-\frac{\beta}{2} \sum_{\alpha=1}^{n}
 \sum_{ij} (J_{ij}-\lambda \delta_{ij}) s_i^{\alpha} 
s_j^{\alpha} }}}
\label{rep3}
\end{equation}
where we have chosen to take the expectation value of the first
replica,
although clearly any other replica will do.

\item {\em Dynamics}:  
The dynamic method \cite{cirano,Sozi}  consists of
calculating the average in 
(\ref{quenched}) by considering the solution of the Langevin equation:
\begin{equation}
\gamma {\dot{s}}_i= -\beta \frac{\partial E_{\boldsymbol{J}}}{\partial s_i} +
\beta h_i + \rho_i
\label{Langevin}
\end{equation}
where the $\rho_i$ are independent Gaussian white noises with variance 
$=2\gamma$.

The energy might be complex: this poses no problem (at least
for linear systems \cite{CLang}). Starting from $t=0$, we are
guaranteed that at long times $t_o$:
\begin{equation}
\langle A({\boldsymbol{s}}) \rangle = 
\lim_{t \rightarrow \infty} \langle  A({\boldsymbol{s}}(t_o))
\rangle_\rho
\label{timeavg}
\end{equation}
where $\langle \bullet \rangle$ denotes thermodynamic average and
$\langle \bullet \rangle_\rho$    average over the process,
i.e. over the noise realisation.
We obtain an expression for the average Green function
(\ref{quenched}) as:
\begin{equation}
{\overline{
G(\lambda)
}}=
\left. i \lim_{t_o \rightarrow \infty}
\sum_k {\overline{ 
\frac{\partial \langle s_k(t_o) \rangle_\rho}{\partial h_k}}} \right|_{h=0} 
\label{response}
\end{equation}
In practice, one calculates the dynamics averaged over both thermal
noise and disorder and in the large $N$ limit, as we shall see below.
\end{itemize}

The problem of treating the denominator is not exclusive of Gaussian
systems, it appears whenever we wish to obtain the correct quenched
averages
over disorder. 
For example,
the  energy (\ref{linear}) can be modified to obtain the standard
spin-glass model:
\begin{equation}
E_{\boldsymbol{J}}^{nl}= i E_{\boldsymbol{J}}(h)  + 
m \sum_i   s_i^2 +  g \sum_i  s_i^4
\label{nonlin}
\end{equation}
and we may wish to calculate averages of any observable
$A({\boldsymbol{s}})$.
As soon as $g > 0$ the  system becomes as
complicated as can be, with all the subtelties of  spin-glasses.
Once we abandon the Gaussian world, 
 the three methods encounter difficulties:

\begin{itemize}

\item
{\em Supersymmetry}: there is no obvious way to write $1/Z_{\boldsymbol{J}}$
in general as an integral over an exponential.
This {\em does not} mean that the supersymmetry method
is entirely inapplicable
for non-Gaussian systems: even though when the energy is not quadratic
this method {\em does not} give the Boltzmann-Gibbs measure,
it can still be useful in some cases, as we shall see below.

\item
{\em Replicas}: In contrast to (\ref{partition-1}),  expressions
(\ref{rep1}) and (\ref{rep2}) are formally valid for non-quadratic
energies. Thus, the replica trick has been applied successfully to
 the study of
many complex  systems, spin-glasses being the main example.   
The expectation values of an observable $A$ can  in general be written
as:
\begin{equation}
{\overline{\langle A \rangle }} 
\sim_{n \rightarrow 0} {\overline{
  \int \Pi_{\alpha=1}^{n} 
{\boldsymbol{ds^\alpha }}
\; A({\boldsymbol{s^{(1)}}})
 \; e^{-\beta \sum_\alpha E_{\boldsymbol{J}}(\boldsymbol{s^\alpha})}
}}
\label{rep4}
\end{equation}

From the point of view of making the results rigorous (or even
reliable), there is the following difficulty:
 a closed analytic expression 
in terms of $n$ can be obtained in some limit, typically large $N$. 
This poses the problem that the limits $N \rightarrow \infty$ and $n 
\rightarrow 0$ may not commute -- 
and indeed in most interesting cases they do not. In those cases
we have to
consider the assumed infinite-$N$ continuation valid around $n=0$ as a guess 
(see however Ref. \cite{KK}).

\item
The {\em dynamic} expression (\ref{timeavg}) shares with the replica
treatment the advantage of being equally valid for linear or nonlinear
problems. There is however  a problem also here: 
(\ref{timeavg}) holds to the extent that we make $t_o\rightarrow
\infty$ before any other limit, in particular $N \rightarrow \infty$.
Again, in many interesting (nonlinear) problems  
these limits do not commute: in physical terms this means that an
infinite system is not able to equilibrate at finite times \cite{Bocukume}.
This is indeed the physical situation one wishes to reproduce in glassy
systems. 
However, one may still be interested in
 knowing what happens in times that diverge with the system size, and in
 particular to reproduce the equilibrium situation - even if it might be
 unreachable in 
a realistic situation \cite{Sozi1}.
To do this, the $N \rightarrow \infty$ solutions must be supplemented 
with activated, `instanton' solutions \cite{Loio}: this problem has not yet been
solved in general.

\end{itemize}

\vspace{.5cm}

{\bf Dynamics is a generalisation of supersymmetry.
}

\vspace{.5cm}

Let us see that the supersymmetry method is 
 a {\em `time-less'} version of  dynamics (\ref{Langevin}). We
 compute
the solutions of the stochastic equation: 
\begin{equation}
0= -\beta \frac{\partial E_{\boldsymbol{J}}}{\partial s_i} + \beta h_i + \rho_i
\label{LL}
\end{equation}
There is no time-dependence, and the $\rho_i$ are Gaussian variables
of variance $2\gamma$.
If $E_{\boldsymbol{J}}(h)$ is quadratic the system (\ref{LL}) has a single solution
\begin{equation}
s_i= -i T \sum_{j} [\lambda {\boldsymbol{I}} - 
{\boldsymbol{J}}]^{-1}_{ij} (\beta h_j+\rho_j)
\end{equation}
Denoting $\langle A({\boldsymbol{s}}) \rangle$ the average of $A$
evaluated
over the ($\rho$-dependent) solutions, we have :
\begin{equation}
{\overline{
G(\lambda)
}}=
\left. i \sum_k
\frac{\partial  {\overline{
\langle
s_k
 \rangle_\rho}}}{\partial h_k} \right|_{h=0} 
\label{response1}
\end{equation}
to be compared with (\ref{response}). To see that this gives back the
supersymmetry method, let us write, for the Gaussian case:
\begin{equation}
\langle
 s_k \rangle_\rho =
\left< \int 
{\boldsymbol{ds}}
 \;  s_k \; \Pi_i  \delta\left(-i\beta \sum_j
(\lambda \delta_{ij} - J_{ij}) s_j +\beta  h_i +\rho_i\right) 
\det  [i\beta (\lambda {\boldsymbol{I}} - {\boldsymbol{J}})] \right>
\label{cosa}
\end{equation}
where the determinant guarantees that the  solution for every
realisation of $\rho$ is counted with
the same weight.
Exponentiating the delta function  as usual
 \cite{delta}:
\begin{eqnarray}
\langle
 s_k \rangle_\rho &=&
\left< \int 
{\boldsymbol{ds}}{\boldsymbol{d\hat{s}}}
 {\boldsymbol{d\eta}}
{\boldsymbol{d\eta^*}}
 \;  s_k \; \right. \times \nonumber \\
& &\left. \exp \left\{ 
\sum_{ij} 
-i \beta  [\lambda \delta_{ij} - J_{ij}] (i {\hat{s}}_i s_j + \eta^*_i \eta_j) 
+i  \sum_i {\hat{s}}_i (\beta h_i +\rho_i) 
\right\} \right> \nonumber \\
 &=&\int 
{\boldsymbol{ds}}{\boldsymbol{\hat{s}}}
 {\boldsymbol{d \eta}}
{\boldsymbol{d \eta^*}}
 \;  s_k \; \times \nonumber \\ 
& &\exp \left\{ 
 \sum_{ij} 
-i \beta [\lambda \delta_{ij} - J_{ij}] (i {\hat{s}}_i s_j + \eta^*_i \eta_j) 
+ i \sum_i \beta  {\hat{s}}_i h_i - \gamma \sum_i  {\hat{s}}_i^2
\right\}
\nonumber \\
\label{cosa1}
\end{eqnarray}
which, using (\ref{response1})  yields:
\begin{eqnarray}
G(\lambda)
 &=& i 
\beta \sum_k \int 
{\boldsymbol{ds}}{\boldsymbol{\hat{ds}}}
 {\boldsymbol{d \eta}}
{\boldsymbol{d \eta^*}}
 \;  s_k  {\hat{s}}_k \; \times \nonumber \\ 
& &\exp \left\{ 
\sum_{ij} 
-i \beta [\lambda \delta_{ij} - J_{ij}] (i {\hat{s}}_i s_j + \eta^*_i \eta_j) 
 - \gamma \sum_i   {\hat{s}}_i^2
\right\}
\label{ttt}
\end{eqnarray}
This is an implementation of supersymmetry
like (\ref{susy}), with two ordinary
$({\boldsymbol{s}},{\boldsymbol{\hat{s}}})$
and two Grassmann $({\boldsymbol{\eta}},{\boldsymbol{{\eta^*}}})$
sets of variables. For $\gamma \rightarrow 0$ it can be taken to the form 
(\ref{susy}) by a rotation in the $({\boldsymbol{s}},{\boldsymbol{\hat{s}}})$.

The conclusion we draw from this exercise is that: {\em i)} Supersymmetry
is just `dynamics without time', which strongly suggests that any
problem solvable with the former is solvable with the latter method.
{\em ii)} Supersymmetry can be extended to treat certain nonlinear problems, 
as we shall now show.

\vspace{.5cm}

{\bf Supersymmetry for nonlinear problems.}

\vspace{.5cm}

Equation (\ref{LL}) is not restricted to linear energy functions. If
(\ref{LL}) is nonlinear, but still has one solution, it can be used  to
calculate the expectation value of any function $A({\boldsymbol{s}})$
in its  root. The generalisation of eqs. (\ref{cosa}) and
(\ref{cosa1}) is:
  \begin{eqnarray}
\langle
 A({\boldsymbol{s}}) \rangle_\rho &=&
 \left< \int d{\boldsymbol{s}} \;  A \; \Pi_i \delta\left( -
\beta \frac{\partial E_{\boldsymbol{J}}}
{\partial s_i} +\rho_i\right) 
\det  \left[
\frac{\partial^2 E_{\boldsymbol{J}}}{\partial s_k \partial s_l}\right] \right>
 \nonumber \\
&=&\int d{\boldsymbol{s}}d{\boldsymbol{\hat{s}}} d{\boldsymbol{\eta}}
d{\boldsymbol{\eta^*}}
 \;  A \; \times \nonumber \\ 
& &\exp \left\{- i\beta \sum_i {\hat{s}}_i \frac{\partial E_{\boldsymbol{J}}}
{\partial s_i}
 + \beta 
\sum_{ij} \eta^*_i \frac{\partial^2 E_{\boldsymbol{J}}}{\partial s_i \partial s_j} \eta_j
 - \gamma \sum_i   {\hat{s}}_i^2
\right\}
\label{cosaa}
\end{eqnarray}
This way of imposing a solution has its origin in the path-integral
treatment of gauge theories \cite{ZJ}, where the fermions are called `ghosts'
\cite{Paso0}.

In many cases of interest the equation (\ref{LL}) has many solutions 
for some realisations of $\rho$. If we wish to add the 
values of the observable in every solution we should take
the absolute value of the determinant in  (\ref{cosaa}). In particular, we
 need to do this if we wish to calculate the average {\em number} of 
solutions. Writing this absolute value as an exponential is possible
\cite{K1}, although it involves introducing new fields.

An interesting situation we shall consider here and in what follows 
is  when we {\em do
  not} take the absolute value. Each solution is then added 
  with the sign of the determinant of the matrix of second
  derivatives \cite{Paso1}.  In particular:
\begin{equation}
\langle 1 \rangle = \sum_{solutions} (-1)^{sign}
\label{Mo}
\end{equation}
which is an  invariant only dependent on the topology of the space of
the ${\boldsymbol{s}}$, and independent of the energy function
$E_{\boldsymbol{J}}$ \cite{K1}.
 For the usual case of the ${\boldsymbol{s}}$ forming a flat space  
and $E_{\boldsymbol{J}}({\boldsymbol{s}}) \rightarrow \infty$ 
as $|{\boldsymbol{s}}|
\rightarrow \infty$
the invariant is {\em one}. 
 In cases in which there are many solutions, the method does not select
 the lowest ones,
 but averages flatly (apart from the sign of the Hessian)
 over {\em all solutions} \cite{foot1}: it is in this sense that supersymmetry 
fails. 

In any case, as mentioned above, 
one is not calculating the Gibbs measure,
but just values over local minima and saddles.
There are however interesting nonlinear problems
having a finite number of solutions for which  
there is no reason to abandon supersymmetry.

One of the most interesting applications involving non-gaussian
problems are the quantum systems.
A note on terminology is necessary  for what follows: In quantum systems,
we can distinguish two ways in which nonlinearity may appear:
in the wavefunction and/or  in the Hamiltonian.
 In the former case, one has a nonlinear Schroedinger equation,
containing for example terms cubic {\em in the wavefunction} (see 
Eq. (\ref{nlins}) below). 
In the latter case,  one generally considers a usual, {\em linear}
 Schroedinger problem, but the Hamiltonian contains terms of degree
 higher than two in the creation and destruction operators. It is then
the path integral that is non-Gaussian, since the  {\em  action} 
is no longer quadratic.
We shall discuss below both cases.

\vspace{.5cm}

{\bf Functional expression for dynamics.}

\vspace{.5cm}

We can see more clearly  the relation between supersymmetry and dynamics
by  constructing a functional
expression for the equation (\ref{Langevin}). We use
exactly
the same procedure as in (\ref{cosaa}), with now delta-functions and
Jacobians promoted to functionals of the trajectories. 
  \begin{eqnarray}
 \langle A({\boldsymbol{s}}(t_o)) \rangle_\rho 
 &=&
\int D{\boldsymbol{s}}
     D{\boldsymbol{\hat{s}}}
     D{\boldsymbol{\eta}}
     D{\boldsymbol{\eta^*}}
 \; A({\boldsymbol{s}}(t_o)) \; \times \nonumber \\ 
 & & \exp  \left\{
- i \beta\sum_i \int dt \; {\hat{s}}_i 
\frac{\partial E_{\boldsymbol{J}}}{\partial s_i}
 +\beta \sum_{ij}\int dt \;
 \eta^*_i \frac{\partial^2 E_{\boldsymbol{J}}}{\partial s_i
 \partial s_j}
 \eta_j \right. \nonumber \\
& & \left. + \gamma \sum_i
\int dt (\eta_i^* {\dot{\eta}}_i- i {\hat{s}}_i {\dot{s}}_i \;- {\hat{s}}_i^2)
\right\}
\label{cosab}
\end{eqnarray}

This functional equation can be viewed either as the
de Dominicis-Janssen, Martin-Siggia-Rose \cite{DJ,MSR} functional
 expression for the Langevin dynamics -- with the determinant
 exponentiated through ghosts -- or as the path-integral expression
 for supersymmetric quantum mechanics \cite{Witten}.

{\em Here we see clearly that  by expressing
expectation values dynamically the
  problem  now becomes, just like in the case of supersymmetry and replicas,
  the computation of an integral of an exponential},
 albeit
a functional one.
This is the usual starting point for the developments in dynamics - at
  least within the physics literature. 
 
This is a good place to see
how one can   calculate with the same method 
the localisation of wavefunctions
in a nonlinear Schroedinger problem \cite{Tr}:
\begin{equation}
i{\dot{\varphi}}_n= -\frac{1}{2}(\varphi_{n-1}+\varphi_{n+1})
 +(\epsilon_n + \Lambda |\varphi_n|^2) \varphi_n
\label{nlins}
\end{equation}
where $\epsilon_n$ is the site  disorder. 
 The system of equations (\ref{nlins}) together with its conjugate  
can have a single solution (this will surely be the case 
if we fix the initial conditions). In fact, (\ref{nlins}) can be
viewed as a (noisless) dynamical equation for the variables ${\varphi_n}$.
We
 can obtain a functional expression
as in (\ref{cosab})
for this case
  introducing  complex Lagrange multiplier
 time-dependent
fields 
${\hat{\phi}}_n (t)$, and Grassmann fields $\eta_n(t)$, $\eta_n^*(t)$.
The normalisation is guaranteed, even if the action is no longer
 quadratic in the $\phi_n(t)$.

\vspace{1in}

\vspace{.5cm}

{\bf Normalisation and symmetries.}

\vspace{.5cm}

We have three expressions for the expectation of an observable:
using supersymmetry (\ref{cosaa}), replicas  (\ref{rep4}) and
dynamics (\ref{cosab}). All three lend themselves to averaging over
the disorder, and have no uncomfortable normalisations.
Indeed, the three expressions yield 
\begin{equation}
\langle 1 \rangle =1
\label{norm}
\end{equation}
but for apparently different reasons:

\begin{itemize}

\item Within the supersymmetric formalism (\ref{norm}) arises because
  around 
each solution the Grassmann and the 
ordinary variables conspire, just as in the Gaussian case, to give
$\pm 1$
 (the sign of the
determinant of the Hessian). Even when there are many solutions, 
these signs add up to {\em one} because
of topological constraints \cite{foot}. 

Now, even if we did not know where the function (\ref{cosaa}) came
from, we could still  see that the expectation value $ \langle 1 \rangle$
does not depend on $E_{\boldsymbol{J}}$ using the fact that the exponent has the {\em two} 
supersymmetries (which indeed give the name to the approach):
\begin{eqnarray}
 \delta{s_i}=\eta_i \;\; ; \;\; \delta {\eta^*}_i= i {\hat{s}}_i
 \nonumber \\
 \delta{s_i}=\eta^*_i \;\; ; \;\; \delta \eta_i=i {\hat{s}}_i
\label{susy1}
\end{eqnarray}

\item Within the replica formalism (\ref{norm}) just expresses the
  fact that we have an  integral to the $n^{th}$ power,
and we let  $n \rightarrow 0$.
Again, if we did not know where (\ref{rep4}) came from, we could show
that $ \langle 1 \rangle=1$ using the fact that the exponent 
is symmetric with respect to replica permutations.

\item In the {\em causal} 
dynamic treatment starting from an initial condition and
letting  the endpoint free, (\ref{norm}) is just a statement of
 probability conservation \cite{ffnn}.
Also in this case we can see directly from the action that   
 $ \langle 1 \rangle=1$, for reasons of symmetry \cite{ZJ}. One has
the following two supersymmetries, which are the
 generalisation of (\ref{susy1}) to the case {\em `with time'}:
\begin{eqnarray}
 \delta{s_i}&=&\eta_i \;\; \, \;\; \delta {\eta^*}_i= i {\hat{s}}_i
 \nonumber \\
 \delta{s_i}&=&\eta^*_i \;\; \, \;\; \delta \eta_i=i {\hat{s}}_i -
 {\dot{s}}_i
 \;\; \, \;\; \delta  {\hat{s}}_i = -i {\dot{\eta}}^*_i  
\label{susy2}
\end{eqnarray}
\end{itemize}
which, together with time-translation invariance, constitute the full
group of symmetry.

\vspace{.5cm}

{\bf A unifying notation.}

\vspace{.5cm}

We have seen that the methods of supersymmetry and dynamics (itself
also possessing a supersymmetry) are closely connected. In fact, we can 
uncover more algebraic correspondences between the three approaches by
using a suitable notation \cite{K1,Saclay}.
This can be done by
 introducing two anticommuting Grassmann variables $\theta\,, \,\bar\theta$:
\begin{equation}
 [\theta\, , \bar\theta]_+ = \theta^2 = \bar \theta^2 = 0
\; \label{2.2} 
\end{equation}
The integrals over these variables are defined as:
\begin{equation}
\int 1 d\theta = \int 1 d \bar \theta = 0 
\;\;\;\;\;\;\;\;\;\;\;\;\;\;\;
\int \theta d\theta = \int \bar \theta d \bar \theta = 1
\; 
 \label{2.3} 
\end{equation}
We can encode the $s_i$, $\eta_i$, $\eta^*_i$ and ${\hat{s}}_i$ in  
a single           {\em superfield}:
\begin{equation}
 \Phi_i
= s_i + \bar \theta \; \eta_i +  \eta_i^* \; \theta + \hat s_i \; 
\bar \theta
\; 
\theta\label{2.4} 
\end{equation}
Using Eqs. (\ref{2.2})-(\ref{2.3})
 and (\ref{2.4}) one obtains,    in terms of the superfields $\Phi_i$   
\begin{eqnarray}
 \langle A \rangle 
&=& 
\int \prod_i \; D[\Phi_i] \; A \;\; \exp \int \; d\alpha \; \left[
\frac{1}{2} \sum_i \Phi_i(\alpha) \Delta(\alpha,\alpha') \Phi_i(\alpha') 
    -  \beta E_{\boldsymbol{J}}(\Phi(\alpha)) \right] \nonumber\\
&~&
\label{functio}  
\end{eqnarray}
where we have denoted  $\alpha \equiv (\theta, \bar\theta)$,
$d\alpha=  d \theta \; d\bar\theta $ and 
\begin{equation}
\Delta =
\Delta^{SUSY}(\alpha,\alpha') \equiv 2 \gamma 
\label{fg}
\end{equation}
independent of $\alpha,\alpha'$.

The dynamics can be encoded in an expression
formally identical to (\ref{functio}), but now the field dependencies
and integration
variables include time: 
 $\alpha \equiv (\theta, \bar\theta,t)$, $\alpha' \equiv
(\theta', \bar\theta',t')$,  
$d\alpha=  d \theta \; d\bar\theta \; dt$ and 
\begin{equation}
\Delta =
\Delta^{Dyn} =
2 \gamma \delta(t-t')+ 
\gamma \delta'(t-t')(\bar \theta-\bar \theta')(\theta+\theta') 
\label{2.9} 
\end{equation}
Finally, the replica expression is again formally (\ref{functio}),
but  with the identification:
\begin{equation}
\Delta^{Replica} =0 \;\;\; ; \;\;\;
\Phi_i(\alpha) \;\; \leftrightarrow s_i^\alpha \;\;\;\;
 ; \;\;\;\; \int \; d\alpha  \;\; \leftrightarrow \sum_{\alpha=1}^{n}
\end{equation}
(The correspondence between supersymmetry and replicas
can be made to hold even for $\gamma \neq 0$ by using
a term $\Delta^{Replica}= \gamma$ which does not affect the final
result.)

In particular, the expectation values $\langle A \rangle$
associated with the calculation
of the Green function (\ref{rep3}),(\ref{response}) and (\ref{ttt})
can be written
 in terms of  $ A $ which,
 in a notation that highlights the analogies, reads:
\begin{equation}
A = i \beta
\sum_a \int d\alpha d \alpha' \; \Phi_a(\alpha) O(\alpha') \Phi_a(\alpha')
\label{AA}
\end{equation}
with the identifications:
\begin{equation}
 O(\alpha)\equiv \delta_{\alpha,1} \;\;\; ; \;\;\;
  O(\alpha)\equiv \delta (\bar \theta)\delta (\theta)  \;\;\; ; \;\;\;
 O(\alpha)\equiv \delta (t-t_o) \delta (\bar \theta)\delta (\theta) 
\end{equation}
for the replica, the supersymmetry and the
dynamic cases, respectively.
We see that the expressions are analogous to one another.

The important point about expression (\ref{functio}) is that, 
apart from the first term
in the exponent, it has  the same form  as the partition function.
This unified notation is useful as a book-keeping device when we have a
diagrammatic expansion \cite{K2,Saclay}, because diagrams on the
three approaches have the same form. Internal lines involve
integrations over the superspace/replica variable, and the effect of
each method is the same due to relations like:
\begin{equation}
\int  \; 1 \; d\alpha
 =0
\end{equation}
valid in all three approaches (in the replica approach as
$n \rightarrow 0$).

\vspace{.5cm}

{\bf The correspondence at work.}

\vspace{.5cm}

\vspace{.5cm}

{\em One-point functions of random matrices.}

\vspace{.5cm}

We now work out the example of the one-point function for the Gaussian
orthogonal ensemble in parallel with replicas and supersymmetry.
(The problem has also been attacked with dynamics \cite{p2}, but we
will not review this here).
The object is not to discuss how both methods can be used in this case
(this has been done in detail long time ago \cite{Edw,rep}), but rather 
to show how the equality of results follows from the formal
correspondence.
  
We use the functional expression (\ref{functio}) with the energy given
by (\ref{linear}), where the $J_{ij}$ are random Gaussian variables
of variance $N^{-1/2}$. Averaging over the ${\boldsymbol{J}}$, and
expressing
everything in terms of the order parameter:
\begin{equation}
Q(\alpha,\alpha')\equiv \frac{1}{N} \sum_i \Phi_i(\alpha) \Phi_i(\alpha') 
\label{order}
\end{equation}
we get, after a few standard steps (which can be borrowed either from
the supersymmetry or from the replica literature):
\begin{eqnarray}
{\overline{G(\lambda)}} &=&  \int d\alpha d \alpha' O(\alpha')
\langle Q(\alpha,\alpha') \rangle \nonumber \\
\langle Q(\alpha,\alpha') \rangle &=&
\int\; D[Q]\; Q(\alpha,\alpha') \times \nonumber \\
 & &\exp \left\{ -\frac{N}{2}\; Tr\; \ln 
[\Delta {\boldsymbol{\delta}} + i \beta \lambda + \beta^2 Q]   
+\frac{N}{4}\;\beta^2\; Tr\; Q^2  \right\} \nonumber \\
\label{expression}
\end{eqnarray}
Here we have used (\ref{AA}). The square and log functions
are to be understood as functions of $Q$ considered as an operator
(i.e. $Q^2(\alpha,\gamma) = \int d \alpha' Q(\alpha,\alpha')
Q(\alpha',\gamma)$, etc) and 
$Tr\;Q\equiv \int d \alpha Q(\alpha,\alpha)$.
The delta function is either the Kroenecker function (in replica
space) or the superspace delta ${\boldsymbol{\delta}}=(\bar \theta-\bar
\theta')(\theta-\theta')$.
 
Expression (\ref{expression}) can be evaluated by saddle point
integration. 
\begin{equation}
Q^{-1}=(\Delta+i\beta \lambda){\boldsymbol{\delta}}+\beta^2 Q 
\label{sp}
\end{equation}
We can now propose for the saddle point value the most general
(replica and super) symmetric  form for $Q$:
\begin{equation}
Q(\alpha,\alpha') = \tilde q {\boldsymbol{\delta}} + q
\label{order1}
\end{equation}
First note that under operator powers and traces (\ref{order1}) behaves exactly
in the same way whether we interpret it as being a replica matrix
($n \rightarrow 0$)  or 
as a function of two superspace variables.
The saddle point equation then becomes:
\begin{eqnarray}
1&=&i \beta \lambda \tilde q + \beta^2 {\tilde{q}}^2 \nonumber \\
0&=& \gamma \tilde q +i \beta \lambda  q +2\beta^2  q 
\end{eqnarray}
Using (\ref{AA}) we have:
\begin{equation}
{\overline{G(\lambda)}} = i \beta
 \int d\alpha d \alpha' \; Q(\alpha,\alpha') O(\alpha')=i\beta \tilde q
\label{AA1}
\end{equation}
 and this yields the semicircle law in the usual way.

The point worth noting here is that there is a close algebraic
relation between  the replica and the supersymmetric approaches.
Indeed, as we shall stress below, all {\em three} approaches are
essentially isomorphic when restricted to a symmetric ansatz.

\pagebreak

\vspace{.5cm}

{\em Quantum systems with interactions.}

\vspace{.5cm}

As a second example, 
let us briefly see how dynamics can be used as an alternative to
replicas in an interacting quantum system.
 Consider the   system of interacting bosons 
in a random potential~\cite{Chna} with
 imaginary-time action:
\begin{multline}
S = \int {d^2}x\,d\tau\:{\psi^*}\left({\partial_\tau}-
\frac{1}{2m}{\nabla^2}-\mu + V(x)\right)\psi\\
+ \int {d^2}x\,{d^d}x'\,d\tau\:
{\psi^*}(x)\psi(x) u(x-x'){\psi^*}(x')\psi(x')
\label{uunn}
\end{multline}
where $u(x-x')$ is the boson interaction and
$V(x)$ is the random potential.
In order to do the correct averaging over disorder, one can use
 the replica trick, thereby obtaining the averaged action:
\begin{multline}
S = \int {d^2}x\,d\tau\:{\psi_\alpha^*}(x,\tau)\left({\partial_\tau}-
\frac{1}{2m}{\nabla^2}-\mu\right){\psi_\alpha}(x,\tau)\\
- \int {d^2}x\,d\tau\,d\tau'\: \frac{1}{2} v_0\:
{\psi_\alpha^*}(x,\tau){\psi_\alpha}(x,\tau)
{\psi_\beta^*}(x,\tau){\psi_\beta}(x,\tau')\\
+ \int {d^2}x\,{d^2}x'\,d\tau\:
{\psi_\alpha^*}(x){\psi_\alpha(x)}
u(x-x'){\psi_\alpha^*}(x'){\psi_\alpha}(x')
\label{ddoo}
\end{multline}
$\alpha=1,2,\ldots,n$ is a replica index.

We can just as well apply a dynamic treatment here. Going back to
(\ref{uunn}), we can consider $x$ and $\tau$ as the site indices,
$\psi(x,\tau)$ and $\psi^*(x,\tau)$ as the dynamic variables, 
and consider their Langevin evolution in an extra (unphysical)
time $t$:
\begin{equation}
\frac{d\psi(x,\tau;t)}{dt} = - \frac{\delta S}{\delta{\psi(x,\tau)}}+
\rho(x,\tau;t) 
\end{equation}
This `stochastic quantisation' strategy can be implemented for
 fermions as well \cite{Pawu}. 
 We can obtain an expression that is formally identical
to (\ref{ddoo}) (up to a term $\Delta$ as
in (\ref{fg})), but now interpreting the fields $\psi$ as
 superfields, functions of both $x,\tau$ and the superspace variable
 $\alpha \equiv \bar \theta,\theta,t$.
 Diagrams for superfields have the same form as the
 replica ones, and one can also study nonperturbative approximations. 

Let us conclude this section by remarking that for this last case
there is another (more physical) approach: the treatment of quantum dynamics
with a thermal bath \`a la Schwinger-Keldysh (see the first
of Refs. \cite{Ka}).
This has the advantage of not having to introduce
an extra time.

\vspace{.5cm}

{\bf Order parameters, symmetry breaking.}

\vspace{.5cm}

Order parameters can be of vector nature $\Psi(\alpha)$, of matrix
nature $Q(\alpha,\alpha')$ and of higher tensorial character.
They may, of course, depend on space.
A special case arises when one wishes to calculate  the two-point
correlation function of random matrices. One needs to introduce
 two sets of superfields, or of replicas
 $\Phi_i^{(1)}(\alpha),\Phi_i^{(2)}(\alpha)$,
 and ends up with an order parameter:
\begin{equation} 
{\boldsymbol{Q}}= \left( 
\begin{array}{cc}
Q^{(11)}  & Q^{(12)} \\
 Q^{(21)}  & Q^{(22)}
\end{array} \right)
\end{equation} 
where $N Q^{(ab)}(\alpha,\alpha') \equiv \sum_i \langle \Phi_i^{(1)}(\alpha)
\Phi_i^{(2)}(\alpha') \rangle$ for $a,b=1,2$.

The different solutions can be classified according to the manner
in which the symmetry is broken.

\begin{itemize}

\item {\em Symmetric} order parameters appear in the solution of Gaussian
one-point problems. This corresponds, as we have seen in the previous section,
to replica-symmetric/supersymmetric solutions. In the dynamic treatment,
the fact that correlation functions  
satisfy supersymmetry (\ref{susy2}) 
is equivalent to stating that {\em
the
system is in equilibrium, and satifies stationarity as well as the 
fluctuation-dissipation theorem}. 
The dynamics of glassy systems in the high temperature phase is of this
kind, and can be solved easily \cite{Sozi} 
in all the cases in which the replica trick
calculation can also be implemented. (For an explicit presentation
of the  algebraic
connection between the two methods, see \cite{K1}).

\item{\em Vector breakings}

Within the replica trick such form of symmetry breaking
appears when the order parameter is a vector in replica space,
and all components are not equal \cite{dopa}.
For matrix order parameters, vector breakings are those such that
the vector $\Psi$ defined as:
\begin{equation}
\Psi(\alpha) \equiv \int d\alpha' Q(\alpha,\alpha')
\label{crt}
\end{equation}
is itself non-symmetric, i.e. dependent on $\alpha$. 
{\em The
same definition can be applied to supersymmetric and dynamic solutions},
with the  substitution of `replica-symmetry' by `super-symmetry'.
There are several examples of such symmetry-breaking fields in the literature:
{\em i)}
vectors in replica space \cite{dopa1} were considered
 in the study of instantons in the random field Ising
 model, their supersymmetric and dynamic counterparts \cite{Cukuunp}
have closely related properties. 
{\em ii)} Replica  matrices with vector type 
were considered \cite{Cagapa} in the computation of 
saddles in free-energy landscapes, and also
in \cite{Kame} for the two-point functions for random matrices.
A related scheme with matrices is the `two block model' \cite{Brmo},
(the first attempt at replica symmetry breaking)
used to count solutions of a spin-glass equations. For this last example
 there is a supersymmetry-breaking ansatz shown to have the same properties
\cite{Brmo,K1},
and more recently a causality-breaking dynamics \cite{Biku}.


\item{\em Matrix breakings}: This appear only for two (or more) indexed
correlations. They can be characterised by the fact that although
$Q(\alpha,\alpha')$ breaks the symmetry,  
the integral $\Psi$ (Eq. (\ref{crt})) is  itself symmetric (independent
of $\alpha$). 
The best known example of matrix breaking is the Parisi ansatz
\cite{Mepavi}
 in replica space.
In the context of dynamics the solution of the long-time out of equilibrium
evolution of the same systems
\cite{Cuku,Bocukume} is of this kind.
Both the Parisi ansatz and the dynamic solution have been generalised to
order parameters of higher tensorial character \cite{tensor,Cuse}.

Whenever the replica trick is feasable, the dynamic treatment is also
possible. They do not yield the same answers if the system
is not ergodic, 
as one corresponds to
the equilibrium situation and the other to the nonequilibrium dynamics.
Only with  the inclusion of all activated (instanton) processes 
will the dynamic solution reproduce all time regimes, and this is 
not yet available in general \cite{Loio,Biku}. 

\end{itemize}

In several of the cases above, 
the equality between the solutions within the different methods 
stems from an  algebraic correspondence, a generalisation 
  of the kind of that we described in the previous section.

\vspace{.5cm}

{\bf Conclusions}

\vspace{.5cm}

 Having a dictionary that allows to translate developments from one
 method to the other, whenever this is possible, can be useful for
 several reasons.
For example, in the field of structural glasses and supercooled liquids,
 arguably  the most important theoretical challenge
 is the inclusion  of solutions
representing the  activated processes responsible for the smearing of
the purely
{\em dynamic}  transition.
Once these solutions are found, one can envisage constructing formally
analogous solutions in replica space, which one might conjecture would be
responsible
for the disapearence of the {\em thermodynamic} (Kauzman) glass transition, or
for a change in its nature.

 From the point of view of mathematical physics, the dynamic method 
seems a promising strategy,
 since everything that is involved is standard  probability
theory and analysis \cite{alice,mathos}.
Indeed, there seems to be no obstacle of principle for the rigorous
derivation
of the solution of out of equilibrium  glass dynamics
\cite{Cuku,Cuku2,Bocukume},
 at least at the mean-field level.

\vspace{1in}

{\bf Aknowledgements}

\vspace{.5cm}
 
I wish to thank C. Chamon, L. Cugliandolo and G. Lozano
for clarifying discussions and suggestions.



\begin{thebibliography}{99}
\bibitem{Sozi1} Indeed, studying the limit $t \rightarrow \infty$
taken before the large-size limit was the  intention in
the seminal papers of Sompolinsky \cite{So}, and Sompolinsky and Zippelius
\cite{Sozi}.
The solution for finite-time separation sector in 
\cite{Sozi} is correct, while the `hopping dynamics' in 
\cite{So} is not, although many elements encountered there survive
in the correct out of equilibrium version (see \cite{Cuku,Cuku2}).
\bibitem{delta} 
A somewhat more proper way to write this is as the $k \rightarrow
\infty$
limit of:
\begin{equation}
 \delta(f(s)) \sim \int \frac{d{\hat{s}}}{2 \pi} e^{ i{\hat{s}}f(s) -  {\hat{s}}^2/2k}
\end{equation}
Changing variables  ${\tilde{s}}= k{\hat{s}}$ the exponent becomes
 $k[i{\tilde{s}}f(s) -  {\tilde{s}}^2/2]$, and the integrals over $s$ and
${\tilde{s}}$ can be evaluated for large $k$ by saddle point in the complex
 plane. The saddles are the zeroes of $f(s)f'(s)$, but only those with
 $f(s)=0$ dominate.
\bibitem{foot1} This is indeed the reason why the derivation of
  dimensional reduction in \cite{Paso0} fails for the random field
  Ising model.
\bibitem{foot} In general this yields a constant independent of the energy.
For the usual case with flat topology and energy going to $+\infty$ 
as ${\boldsymbol{|s|}}\rightarrow \infty$ the constant is one,
this being  the generalisation of 
the identity 
$(number\; of\; minima) - (number\; of\; maxima)=1$
for functions of  one variable.
\bibitem{ffnn} We can also get  
(\ref{norm}) by summing over all periodic trajectories, in this case
(\ref{norm})
is again a consequence of topology \cite{Witten}.
\end{thebibliography}
\end{document}